\documentclass[12pt]{iopart}

\usepackage{graphicx,epsf, epsfig}
\usepackage[usenames,dvipsnames]{xcolor}
\usepackage[colorlinks=true,linktocpage=false,citecolor=blue,linkcolor=blue,urlcolor=blue]{hyperref}
\usepackage{amsfonts,amssymb,mathrsfs}
\usepackage{multirow}
\usepackage{booktabs,array}
\usepackage{color}
\usepackage{xspace}
\usepackage[nolist]{acronym}

\newcommand{\bbh}{\ac{BBH}\xspace}
\newcommand{\bw}{{\tt BayesWave}\xspace}
\newcommand{\Msun}{\ensuremath{\mathrm{M}_\odot}\xspace}

\begin{document}

\title{Bayesian Inference Analysis of Unmodelled Gravitational-Wave Transients}

\author{Francesco Pannarale}
\ead{francesco.pannarale@ligo.org}
\address{Gravity Exploration Institute, School of Physics and Astronomy, Cardiff University, The Parade,
  Cardiff CF24 3AA, UK}
\address{Dipartimento di Fisica, Universit\`a di Roma ``Sapienza'' \& Sezione INFN Roma1, P.A.\,Moro 5, 00185, Roma, Italy}

\author{Ronaldas Macas}
\ead{ronaldas.macas@ligo.org}
\address{Gravity Exploration Institute, School of Physics and Astronomy, Cardiff University, The Parade,
  Cardiff CF24 3AA, UK}

\author{Patrick J.\,Sutton}
\ead{patrick.sutton@ligo.org}
\address{Gravity Exploration Institute, School of Physics and Astronomy, Cardiff University, The Parade,
  Cardiff CF24 3AA, UK}

\date{\today}

\begin{abstract}
  We report the results of an in-depth analysis of the parameter
  estimation capabilities of {\bw}, an algorithm for the
  reconstruction of gravitational-wave signals without reference to a
  specific signal model.  Using binary black hole signals, we compare
  {\bw}'s performance to
  the theoretical best achievable performance in three key areas: sky
  localisation accuracy, signal/noise discrimination, and waveform
  reconstruction accuracy.  {\bw} is most effective for signals that
  have very compact time-frequency representations. For binaries,
  where the signal time-frequency volume decreases with mass, we find
  that {\bw}'s performance reaches or approaches theoretical optimal
  limits for system masses above approximately 50 $M_\odot$.
  For such systems \bw is able to localise the source on the sky as
  well as templated Bayesian analyses that rely on a precise signal
  model, and it is better than timing-only triangulation in all cases.
  We also show that the discrimination of signals against glitches and
  noise closely follow analytical predictions, and that only a small
  fraction of signals are discarded as glitches at a false alarm rate
  of 1/100\,y.  Finally, the match between \bw-reconstructed signals
  and injected signals is broadly consistent with first-principles
  estimates of the maximum possible accuracy,
  peaking at about $0.95$ for high mass systems and decreasing for
  lower-mass systems.  These results demonstrate the potential of
  unmodelled signal reconstruction techniques for gravitational-wave
  astronomy.
\end{abstract}

\maketitle 

\begin{acronym}
\acrodef{BH}[BH]{black hole}
\acrodef{BBH}[BBH]{binary black hole}
\acrodef{cWB}[{\tt cWB}]{coherent WaveBurst}
\acrodef{FAR}[FAR]{false alarm rate}
\acrodef{GW}[GW]{gravitational wave}
\acrodef{Oone}[O1]{first Advanced LIGO Observing Run}
\acrodef{SNR}[SNR]{signal-to-noise ratio}
\end{acronym}

\section{Introduction}
\label{sec:intro}
The recent direct detection of gravitational waves by Advanced LIGO
and Advanced Virgo has opened a new window in observational astronomy.
On 14 September 2015, LIGO made the first ever observation of a
\ac{BBH} merger~\cite{GW150914}.  The signal, denoted GW150914, has
been followed by other detections of \ac{BBH} mergers ---
GW151226~\cite{GW151226}, GW170104~\cite{GW170104},
GW170608~\cite{GW170608}, and GW170814~\cite{GW170814} --- and by the
detection of a binary neutron star inspiral, GW170817~\cite{GW170817}.
The \ac{BBH} observations have revealed the existence of a previously
unknown population of high-mass ($> 20\,M_\odot$) \acp{BH}, with
implications for our understanding of stellar
evolution~\cite{GW150914astro, Belczynski:2016obo, Belczynski:2017gds,
  Stevenson:2017tfq}, and have allowed the first tests of general
relativity in the dynamical, strong-field regime
\cite{GW150914testingGR, GW151226, GW170104}.

The interpretation of these signals has relied upon precise signal
models (``templates'') \cite{PE-GW150914} that can be compared to the
data.
However, there are many possible emission mechanisms beyond \acp{BBH}
for which the \ac{GW} radiation cannot be easily modelled, such as
core-collapse supernov\ae~\cite{Ott:2008wt, Kotake:2012iv,
  Yakunin:2017tus, Richers:2017joj, Kuroda:2016bjd}, post-merger
emission by hypermassive neutron stars in binary neutron star
mergers~\cite{Bauswein:2011tp, Hotokezaka:2013iia, Kastaun:2014fna,
  Takami:2014tva, Bernuzzi:2015opx, Bernuzzi:2015rla,
  Rezzolla:2016nxn, Shibata:2017xht}, and magnetar
flares~\cite{1995MNRAS.275..255T, 2001MNRAS.327..639I,
  2011PhRvD..83j4014C}, for which matched filtering is not applicable.
This has spurred the development of tools for the detection and
characterisation of \emph{generic} \ac{GW} transients (a.k.a.,
``bursts'')~\cite{cWB1,cWB2,oLIB,STAMPASStandardRef,X1,X2}.  Among
them is \bw~\cite{Cornish:2014kda}, a Bayesian parameter estimation
algorithm for the reconstruction of generic \ac{GW} transients.
Instead of relying on a precise signal model, \bw fits linear
combinations of basis functions to the data in a manner consistent
with either a \ac{GW} or background noise artifacts (``glitches'').
Given a potential \ac{GW} trigger, \bw performs a Bayesian analysis
under the signal and the glitch hypotheses, reconstructs the
gravitational waveform, and provides estimates of model-independent
parameters, such as the signal duration, bandwidth, and sky location.
This tool was used successfully, for example, in the follow-up of
GW150914 and GW170104~\cite{GW150914, TheLIGOScientific:2016uux,
  GW170104}.

In this paper, we subject \bw to a series of tests in order to
validate it and assess its performance against first-principles
estimates.  While there have been some studies of the performance of
such algorithms for various kinds of burst signals (see for
example~\cite{Klimenko:2011hz,Abbott:2011ys,Essick:2014wwa,Becsy:2016ofp}),
very little has been done to compare their performance to
first-principles expectations; \textit{i.e.}, we do not know if
presently available tools are performing close to optimally, or if
there is significant room for improvement.  We address this by
assessing \bw in three critical areas:
\begin{enumerate}
\item \textit{Sky localisation:} How accurately can \bw determine the
  direction to the \ac{GW} source, compared to ideal matched-filtering
  algorithms?
\item \textit{Signal--glitch discrimination:} How robustly can \bw
  distinguish true \ac{GW} signals from non-Gaussian background noise
  artifacts?
\item \textit{Waveform reconstruction:} How does the accuracy of
  {\bw}'s reconstructed gravitational waveforms compare to first
  principles estimates of the possible accuracy of unmodelled
  reconstructions?
\end{enumerate}
We answer these questions by applying \bw to a set of simulated
\ac{BBH} signals added to simulated Advanced LIGO and Advanced Virgo
data \cite{Hooper:2011rb}.  While accurate templates are available for
\ac{BBH} signals, \bw does not use this information.  Using \ac{BBH}
templated signals for our tests allows us to compare the performance
of \bw to the case of ideal matched-filtering, which does rely on a
precise signal model.
Despite not using a signal model, we find that the performance of \bw
is remarkably close to optimal in most cases, and we note the
conditions under which performance is less than optimal.

The paper is organised as follows.  In Sec.\,\ref{sec:BW}, we briefly
discuss the \bw algorithm.  In Sec.\,\ref{sec:results} we describe the
test performed to assess the performance of \bw and discuss our
results.  Finally, our conclusions are summarized in
Sec.\,\ref{sec:conclusions}.

\section{BayesWave}
\label{sec:BW}
\bw is a Bayesian follow-up pipeline for \ac{GW} triggers.  It is
designed to distinguish \ac{GW} signals from non-stationary,
non-Gaussian noise transients (\textit{i.e.}, glitches) in
interferometric \ac{GW} detector data, and to characterize the signals
themselves~\cite{Cornish:2014kda, Littenberg:2014oda}.  \bw uses a
multi-component, parametric noise model of variable dimension that
accounts for instrument glitches.  These are modeled using a linear
combination of Morlet-Gabor continuous wavelets.  A trans-dimensional
reversible jump Markov chain Monte Carlo algorithm allows for the
number of wavelets to vary and to explore the parameters of each
wavelet.  \ac{GW} transients of astrophysical origin are
(independently) modelled with the same technique: a single \ac{GW}
signal model is built at the center of the Earth and projected onto
each detector in the network, taking into account the response of the
instrument and the source sky-location, which feeds two parameters
(\textit{i.e.}, right ascension and declination) into the
reconstruction effort.  A linear combination of wavelets constituting
a glitch model is instead built for each individual detector.  In
other words, there is a requirement for signals to appear coherently
in the data, but not for glitches.

The \bw algorithm compares the following hypotheses: (1) the data
contain only Gaussian noise, (2) the data contain Gaussian noise and
glitches, and (3) the data contain Gaussian noise and a \ac{GW}
signal.  The comparison is performed in terms of the marginalized
posterior (evidence) for each hypothesis.  When testing the signal
hypothesis, \bw provides a waveform reconstruction, and posterior
distributions for the source sky location parameters and signal
characteristics, such as duration, bandwidth, energy, central
frequency.  These may be used to compare the data to theoretical
models and to assess the performance of the pipeline.

\bw has been used in a number of studies so far.  Notably, it was used
as a follow-up analysis to candidate and background events found by
the \ac{cWB} pipeline~\cite{cWB1, cWB2} and matched-filter searches
during the first two Advanced LIGO observing
runs~\cite{TheLIGOScientific:2016uux, Abbott:2017vtc,
  Abbott:2017oio}.  \bw localized the source of the GW150914 event in
a $101$ square degree region with 50\% confidence and set a \ac{FAR}
of $1$ in $67400$ years.  Further, \cite{TheLIGOScientific:2016uux}
tested the ability of \bw in recovering simulated \bbh signals for
sources similar to GW150914.  The match between the reconstructed and
injected waveforms was found to be vary between 90\% and 95\% for
systems with total mass between $\sim 60$\,\Msun and $\sim
100$\,\Msun, and an injected network \ac{SNR} of $20$.  The
sensitivity range, which is tightly correlated to the total mass and
the effective spin of the system, was found to be in the
$400$--$800$\,Mpc interval.  In general, the combined \ac{cWB}-\bw
data analysis pipeline was shown to allow for detections across a
range of waveform morphologies~\cite{Kanner:2015xua,
  Littenberg:2015kpb}, with confidence increasing with the waveform
complexity (at a fixed \ac{SNR}).  This is the case because glitches
can be confused more easily with simple, short \ac{GW} transients,
rather than with complex waveforms in coherent data.  Finally, a
recent study~\cite{Becsy:2016ofp} shows that the two-detector
Advanced LIGO network will be able to achieve an $85$\% and $95$\%
match for \ac{GW} signals with network \ac{SNR} below $\sim 20$ and
$\sim 50$, respectively.  In the same study, the median searched area
and the median angular offset for \bbh sources with total mass between
$30\,M_\odot$ and $50\,M_\odot$ were found to be $99.2$ square degrees
and $25.1$ degrees, respectively.
 
\section{Procedure and Results}
\label{sec:results}

\subsection{Simulated Signal Population}
The source population we choose for our study consists of non-spinning
merging \acp{BBH}.  The values of the individual \ac{BH} masses that
we select are $5\,M_\odot$, $10\,M_\odot$, $50\,M_\odot$, and
$100\,M_\odot$. We consider all 10 possible mass combinations:
$(5,5)\,M_\odot$, $(5,10)\,M_\odot$, $(5,50)\,M_\odot$,
$(5,100)\,M_\odot$, $(10,10)\,M_\odot$, $(10,50)\,M_\odot$,
$(10,100)\,M_\odot$, $(50,50)\,M_\odot$, $(50,100)\,M_\odot$, and
$(100,100)\,M_\odot$.  This population is convenient for a number of
reasons.
\begin{enumerate}
\item The majority of \ac{GW} signals detected by LIGO to date were
  emitted by \ac{BBH} sources~\cite{GW150914, GW151226, GW170104,
    GW170608, GW170814}, and \ac{BBH} mergers are expected to dominate
  the population of \acp{GW} that we detect with second-generation
  instruments~\cite{Dominik:2014yma}.  The \ac{BH} masses of the
  sources detected so far (both the binary constituents and the merger
  remnants) are all encompassed by our choice of parameter space.
\item Accurate and computationally tractable waveform models exist for
  these signals, allowing us to compare the \bw performance to that of
  optimal (template-based) algorithms as reported in the literature.
  Specifically, we use the so-called IMRPhenomB
  approximant~\cite{Ajith:2009bn}.
\item \bw may be able to resolve aspects of the waveform that are not
  included in current templated analyses, such as precessing spins or
  eccentricity.  Ultimately it will be useful to characterise \bw for
  the entire family of \ac{BBH} signals: in this sense, our
  non-spinning study is a first step in this direction.
\item The \ac{SNR} of signals from high-mass systems is concentrated
  in a small time-frequency volume, while the \ac{SNR} of signals from
  low-mass systems, \textit{e.g.}, binary neutron
  stars~\cite{GW170817}, is spread over a much larger time-frequency
  volume. This allows us to probe the performance of \bw relative to
  templated algorithms as a function of the signal time-frequency
  volume, which along with the \ac{SNR} is the key characteristic of a
  signal for burst detection algorithms \cite{rule_of_thumb}.
\end{enumerate}

For each of the 10 mass pairs we generate 20 signals, for a total of
200 simulations, with random sky position, inclination, and
polarisation angle. The distances are selected randomly such that the
coherent network \ac{SNR} is in the range 10--35; \textit{i.e.}, we
use realistic amplitudes for detectable signals. The signals
are added to simulated data for the LIGO-Virgo network H1-L1-V1, which
consists of Gaussian noise following the power spectral density model
of \cite{Hooper:2011rb}. As the \acp{BH} inspiral, the frequency of
the \ac{GW} signal increases until the two bodies merge and the
\ac{GW} emission cuts off.  The merger frequency scales inversely with
system mass, so signals from low-mass systems span the full LIGO/Virgo
sensitive band and therefore have large effective bandwidth and
time-frequency volume. For high-mass systems the effective bandwidth
is much smaller and the signal is concentrated in a relatively small
time-frequency volume. These will have implications for localisation
accuracy and waveform reconstruction that are discussed later in the
text.

For each simulation we analyse 4\,s of data centred on the binary
coalescence time, generated at a sampling rate of
1024\,Hz. 
This data is fed into \bw for analysis.  \bw reports the log evidence
for signal vs.~glitch and for signal vs.~noise hypotheses, a sky map,
reconstructed time-domain waveforms, spectrograms, and estimates of
other properties such as duration, bandwidth, and the \ac{SNR}
recovered in each detector.  In the following subsections we focus on
\bw's performance on spectrograms, signal vs.~glitch discrimination,
and the accuracy of sky localisation and waveform reconstruction.

\subsection{Time-Frequency Signal Content}
\label{sec:tf-content}

As shown in \cite{Littenberg:2015kpb}, the number of wavelets used by
\bw to reconstruct a \ac{GW} signal increases approximately linearly
with \ac{SNR}, at a rate that depends on the signal morphology (higher
for more complex waveforms).  This is consistent with the behaviour
seen in our simulations.  For the \acp{SNR} considered in our study,
the average reconstructed \ac{SNR} per wavelet is typically 5--10.

\begin{figure*}[htb]
  \begin{tabular*}{\textwidth}{c@{\extracolsep{\fill}}c}
    \includegraphics[width=0.5\textwidth,clip=true]{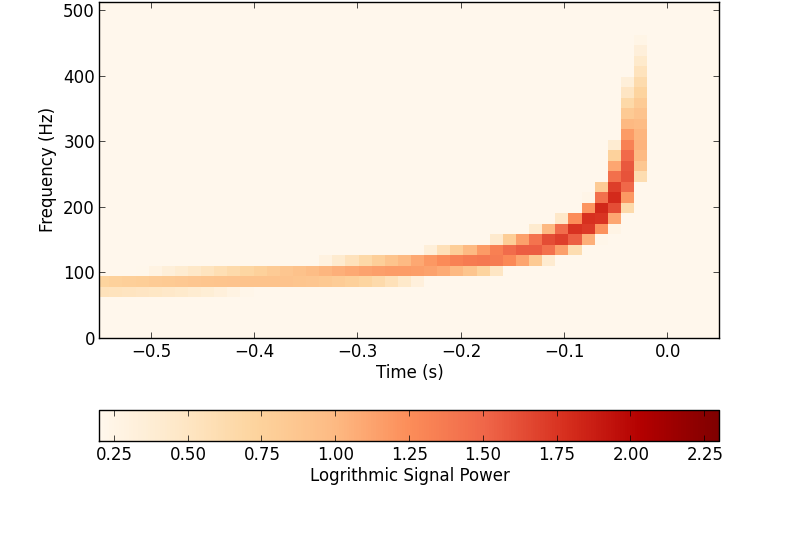}
    \includegraphics[width=0.5\textwidth,clip=true]{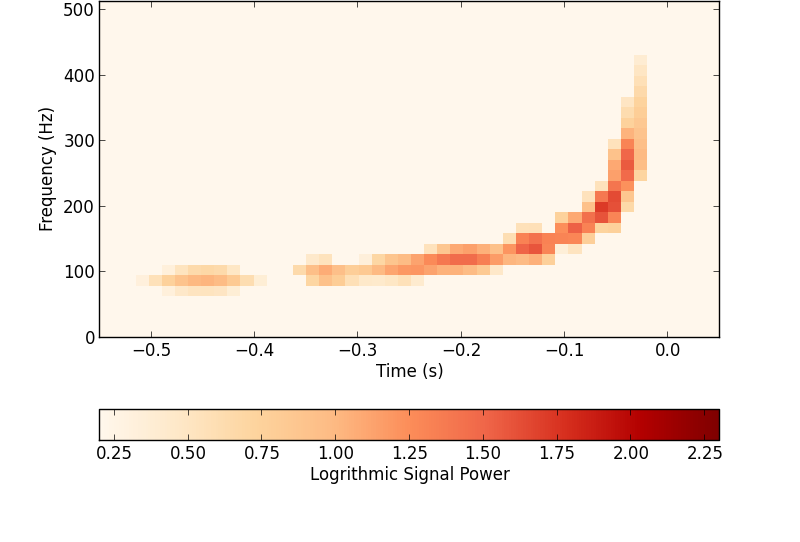} \\
    \includegraphics[width=0.5\textwidth,clip=true]{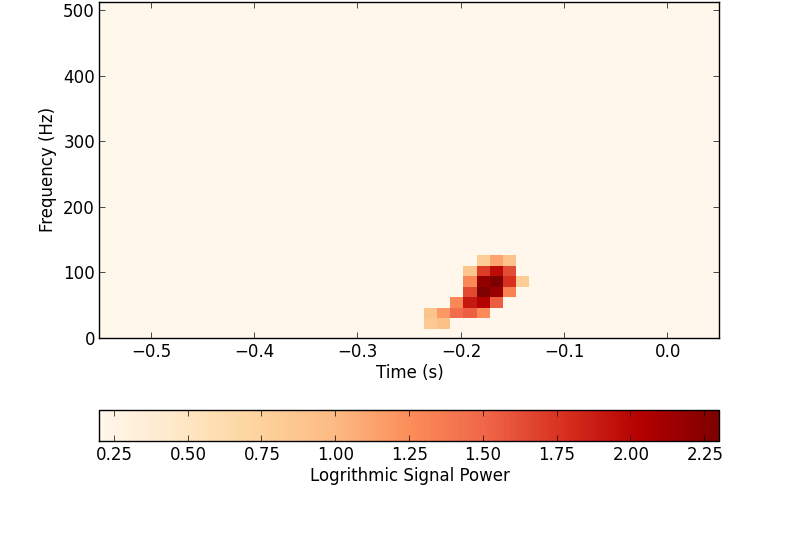}
    \includegraphics[width=0.5\textwidth,clip=true]{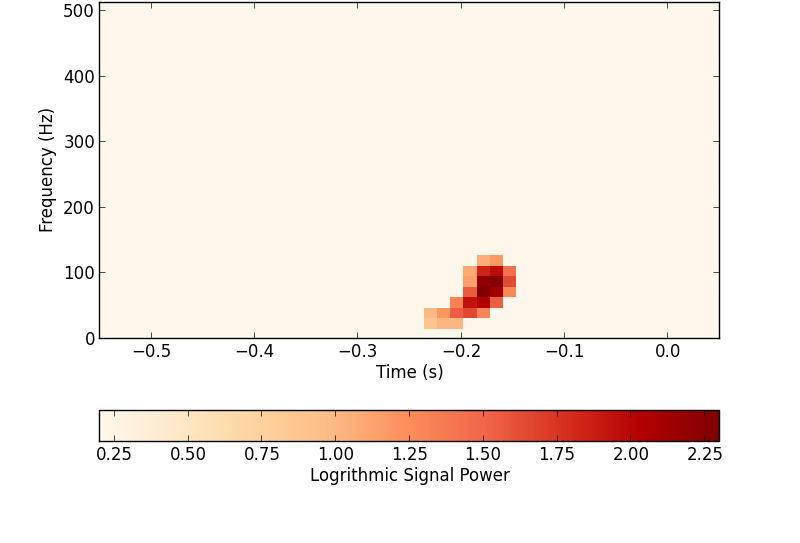}
  \end{tabular*}
  \caption{\label{fig:tfmaps} Whitened
    spectrograms for simulated and recovered signals.  Top:
    (5,5)\,$M_\odot$ simulated (left) and recovered (right)
    signal. The \ac{SNR} per time-frequency pixel is lowest at early times and low
    frequencies; \bw only recovers fragments of this portion of the
    signal.  Bottom: (100,100)\,$M_\odot$ simulated (left) and
    recovered (right) signal. The \ac{SNR} is concentrated into a
    small number of time-frequency pixels which are easily recovered by \bw.
  }
\end{figure*}

\begin{figure*}[htb]
  \begin{tabular*}{\textwidth}{c@{\extracolsep{\fill}}c}
    \includegraphics[width=0.46\textwidth,clip=true]{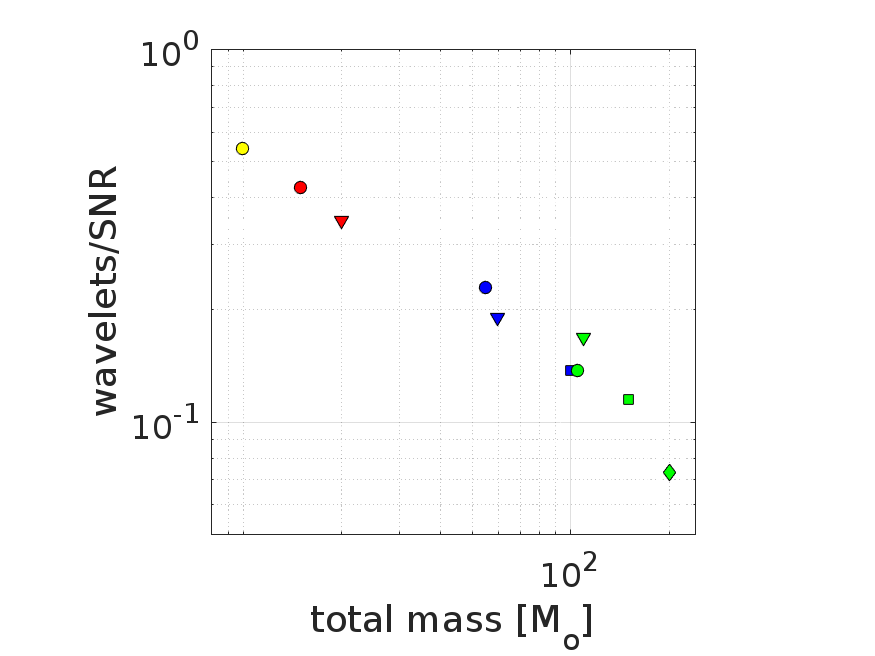}
    \includegraphics[width=0.495\textwidth,clip=true]{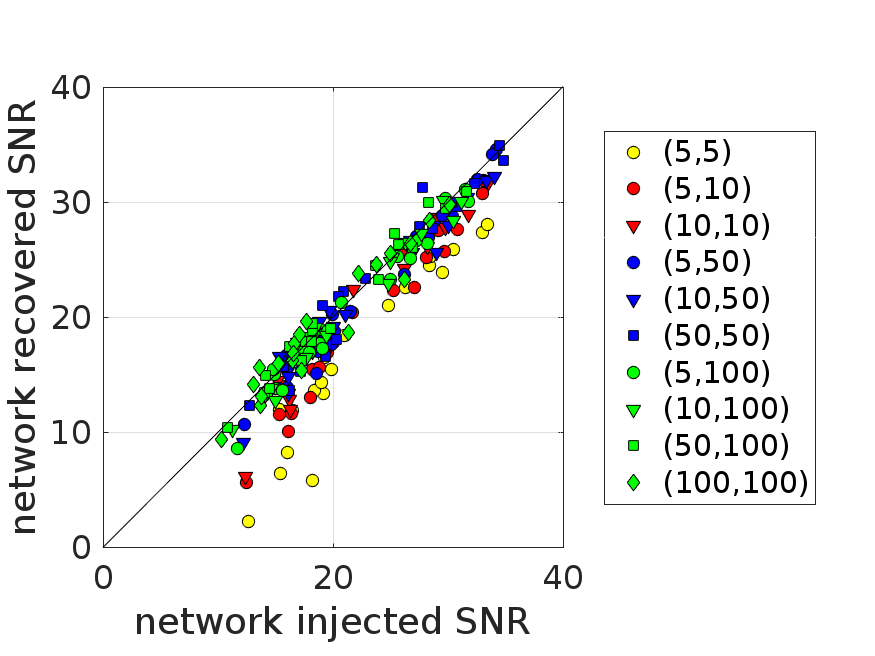}
  \end{tabular*}
  \caption{\label{fig:wavelets} Left: Mean number of wavelets per unit
    injected network \ac{SNR} vs.~system total mass.  As the system
    mass increases, the signal \ac{SNR} in concentrated into a smaller
    time-frequency volume, and can be reconstructed with fewer
    wavelets.  Right: Recovered network \ac{SNR} vs.~injected network
    \ac{SNR}. {\bw} is able to recover the full \ac{SNR} of high-mass
    systems, which occupy a small time-frequency volume.  It
    systematically underestimates the \ac{SNR} of low-mass systems,
    which occupy a larger time-frequency volume at a given \ac{SNR}.}
\end{figure*}

For inspiralling \acp{BBH}, the frequency increases until the two
bodies merge and the gravitational-wave emission cuts off, as the
remnant \ac{BH} rings down. The merger frequency scales inversely with
system mass: low-mass systems produce \ac{GW} signals that have larger
effective bandwidth and time-frequency volume than high-mass systems.
Furthermore, the rate of frequency increase in the signal
(``chirping'') increases with the system mass, so that high-mass
systems have a much shorter duration in the detector sensitive
band. Together these have an important consequence for burst
algorithms such as \bw that rely on time-frequency decompositions:
signals from low-mass systems are spread over a larger time-frequency
area than signals from high-mass systems.  Figure \ref{fig:tfmaps}
shows example spectrograms of the simulated and recovered signals for
the lowest- and highest-mass systems tested.  The low-mass,
$(5,5)\,M_\odot$, simulated signal shown on the top left panel
occupies a time-frequency area greater than the high-mass,
$(100,100)\,M_\odot$, simulated signal shown in the bottom left panel.
As a result, 
\bw is able to recover all of the \ac{SNR} of the high-mass signal
(bottom right panel), but not all of the \ac{SNR} of the low-mass
signal (top right panel).  Figure \ref{fig:wavelets} confirms that the
\ac{SNR} is spread across a larger number of pixels as the system mass
decreases.  Generally, diluting a given total \ac{SNR} among a larger
number of pixels makes it more difficult for {\bw} to reconstruct the
low-\ac{SNR} portions of the signal. This typically results in a lower
reconstructed \ac{SNR}, duration, and bandwidth, which in turn lowers
the accuracy of the sky localisation, signal classification and
waveform reconstruction.

\subsection{Sky Localisation}

There are numerous empirical studies of the sky localisation
capabilities of existing \ac{GW} transient detection algorithms, 
particularly in the context of second-generation \ac{GW}
detector networks (see 
e.g.~\cite{Klimenko:2011hz,Abbott:2011ys,Essick:2014wwa,Becsy:2016ofp}).
The theoretical basis for sky localisation accuracy
is best established for matched-filter searches for binary
coalescences.  As shown by Fairhurst \cite{Fairhurst:2009tc,
  Fairhurst:2010is, Fairhurst:2012tf, Fairhurst:2017mvj}, the localisation is
based primarily on triangulation via the time-of-arrival differences
between the detector sites. The one-sigma measurement uncertainty in
the time of arrival is given by
\begin{equation}
  \sigma = \frac{1}{2\pi\rho\,\sigma_f} \, ,
\end{equation}
where $\rho$ is the matched-filter \ac{SNR} in the detector and
$\sigma_f$ is the effective bandwidth of the signal; see
\cite{Fairhurst:2010is} for definitions.  Ignoring the phase and
amplitude measured in each detector, Fairhurst shows that one can
construct a localisation matrix that defines the contours of fixed
probability,
\begin{equation}
  \mathbf{M} = \frac{1}{\sum_k \sigma_k^{-2}} \sum_{i,j}
  \frac{(\mathbf{d}_i - \mathbf{d}_j) (\mathbf{d}_i -
    \mathbf{d}_j)^T}{2\sigma_i^2\sigma_j^2} \, ,
\end{equation}
where $\mathbf{d}_i$ is the position vector of detector $i$ and
$\sigma_i$ is the timing uncertainty in that detector.
The expected sky localisation accuracy with containment probability
$p$ is given by
\begin{equation}\label{eq:area}
  A(p, \mathbf{r}) = 2\pi \sigma_1 \sigma_2 [- \log(1-p)] \, ,
\end{equation}
where $\sigma_1, \sigma_2$ are the inverse square roots of the
eigenvalues of the matrix $\mathbf{M}$ after it has been projected
onto the sky in the direction $\mathbf{r}$,
\begin{equation}
  \mathbf{M}(\mathbf{r}) = \mathbf{P}(\mathbf{r}) \, \mathbf{M} \,
  \mathbf{P}(\mathbf{r}) \, , \quad \mathbf{P}(\mathbf{r})= \mathbf{I} - \mathbf{r} \mathbf{r}^T \, ,
\end{equation}
where $\mathbf{I}$ is the identity matrix.  Since the approximation
(\ref{eq:area}) ignores the phase and amplitude information, it can be
considered as a worse-case estimate of the localisation capability.

As shown in \cite{Grover:2013sha} and \cite{Fairhurst:2017mvj}, requiring a
consistent signal phase and polarisation between the detectors
improves the localisation accuracy by an amount which can be
approximated by using a timing uncertainty of
\begin{equation}\label{eq:coherent}
  \sigma_t^{c} =
  \frac{1}{2\pi\rho\,\sigma_f}\left(\frac{\sigma_f^2}{\bar{f^2}}\right)^{1/4}
  \, ,
\end{equation}
where $\overline{f^2}$ is the second frequency moment of the signal.
Since $\overline{f^2} > \sigma_f^2$, Eq.\,(\ref{eq:coherent}) will
yield smaller localisation areas.  In their study of binary inspiral
signals of total mass up to 20\,$M_\odot$, Grover \textit{et
  al.}~demonstrated that this phase and polarisation correction
reduces the predicted localisation areas by a factor of 2--3 relative
to timing alone.  Finally, Grover \textit{et al.}~also demonstrated
that a full Bayesian analysis using signal templates achieves sky
localisation accuracies that are still better by a median factor of
1.6; we take this Bayesian analysis to represent the ``best possible''
performance in the case where the signal waveform is known.
\cite{Berry:2014jja} found a similar result for binary neutron star
sources, with a median factor of $\sim 1.3$ for the 50\% localisation
area.

\begin{figure*}[htb]
    \includegraphics[width=\textwidth,clip=true]{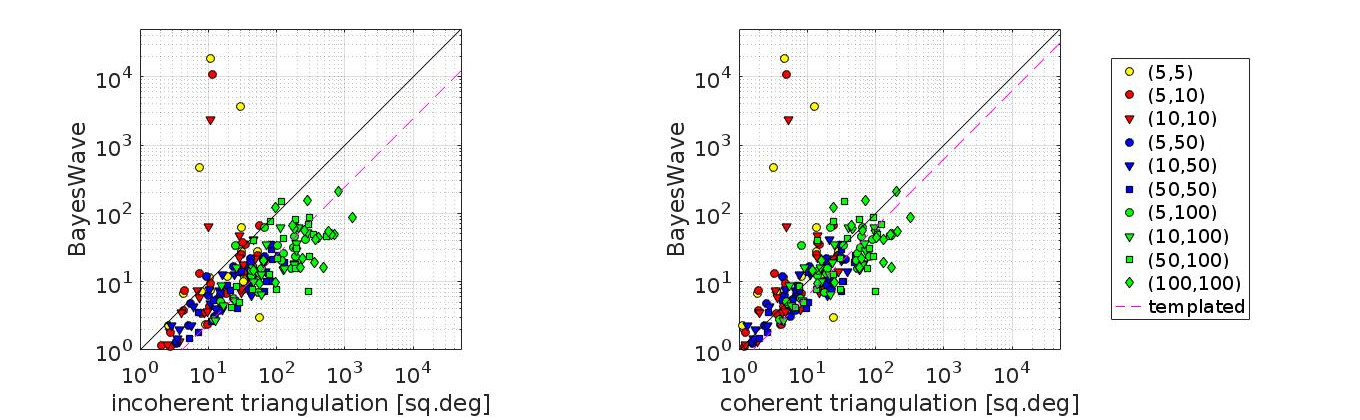}
  \caption{\label{Fig:SkyLoc1} 50\% containment localisation areas
    measured by {\bw} versus those predicted by timing-only
    ``incoherent'' triangulation (left) and phase- and
    polarisation-corrected ``coherent'' triangulation (right). The
    dashed line indicates the approximate median performance of a
    templated Bayesian analysis as reported in~\cite{Grover:2013sha}.
    \bw systematically outperforms the timing-only predictions for all
    mass pairs.  It also systematically outperforms the predictions of
    phase- and polarisation-corrected triangulation for all but the
    lowest-mass systems, despite not using a signal template.  For
    system masses above 50\,M$_\odot$ the \bw performance is
    approximately equal to that of the templated Bayesian analysis.
    In both cases smaller-bandwidth signals tend to have larger
    localisation areas, as expected. 
    The small number of outliers are signals from low-mass systems 
    that \bw was unable to reconstruct accurately.}
\end{figure*}

\begin{figure*}[htb]
    \includegraphics[width=\textwidth,clip=true]{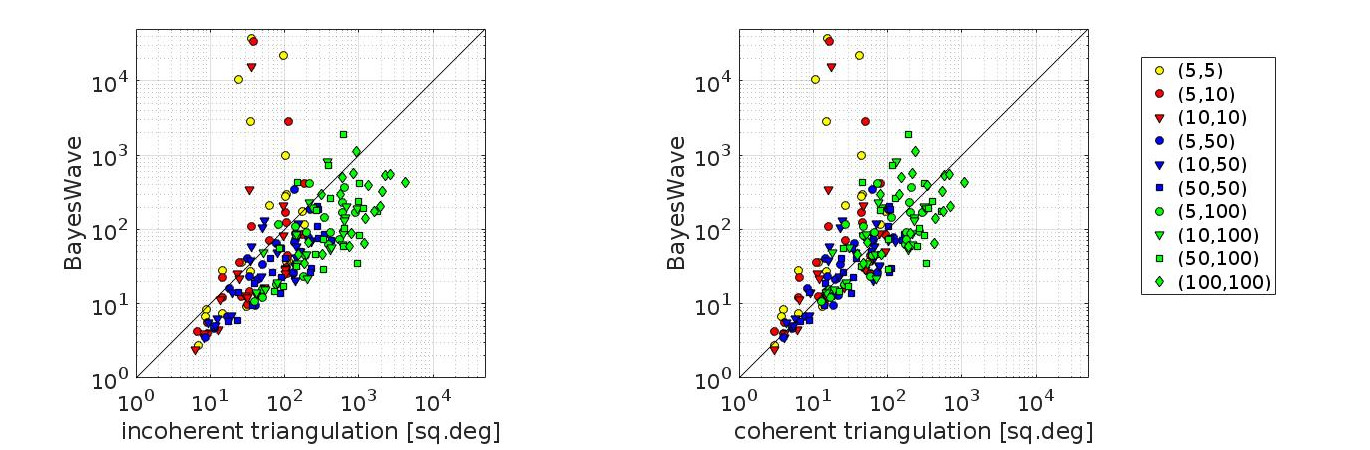}
  \caption{\label{Fig:SkyLoc2} 90\% containment localisation areas
    measured by {\bw} versus those predicted by timing-only
    ``incoherent'' triangulation (left) and phase- and
    polarisation-corrected ``coherent'' triangulation (right).  \bw
    systematically outperforms the timing-only predictions for all
    mass pairs.  It also systematically outperforms the predictions of
    phase- and polarisation-corrected triangulation for all but the
    lowest-mass systems, despite not using a signal template.  In both
    cases smaller-bandwidth signals tend to have larger localisation
    areas, as expected.  
    The small number of outliers are signals from low-mass systems 
    that \bw was unable to reconstruct accurately.}    
\end{figure*}

Table~\ref{tab:ratios} and Figs.\,\ref{Fig:SkyLoc1}
and~\ref{Fig:SkyLoc2} compare the 50\% and 90\% localisation areas
reported by \bw to the predictions of timing-only and phase-corrected
triangulation. We see that \bw easily outperforms the timing-only
predictions. It also outperforms the predictions of phase- and
polarisation-corrected triangulation for all but the lowest-mass
systems, despite not using a signal template. Indeed, for system
masses above 50\,M$_\odot$ the \bw performance is approximately equal
to that of the optimal templated Bayesian analysis reported in Grover
\textit{et al.}~\cite{Grover:2013sha}.

We conclude that {\bw} is able to localise a gravitational-wave source
on the sky as well as a templated analysis despite not using signal
templates, provided the signal \ac{SNR} is concentrated in a
sufficiently small time-frequency volume ($\lesssim 10$ wavelets).
Furthermore {\bw} still performs reasonably well --- within a factor
of 2 in area --- for higher time-frequency volume signals even for
large containment regions.

\begin{table}[!tb]
  \caption{\label{tab:ratios} Median ratio of 50\% and 90\% sky localisation areas reported by {\bw} to those predicted by triangulation. {\bw} typically outperforms the predictions of incoherent (timing-only) triangulation in almost all cases, and outperforms the predictions of coherent (phase- and polarisation-corrected) triangulation for systems of total mass above 50\,M$_\odot$. For comparison, \cite{Grover:2013sha} report that the 50\% localisations from an optimal templated Bayesian analysis are typically $\sim$0.6-0.7 of those of coherent triangulation [see also \cite{Berry:2014jja}]; we see that {\bw} performs comparably for system masses around 100 M$_\odot$ or more.}
  \resizebox{\columnwidth}{!}{
    \begin{tabular}{c@{\hspace{0.3cm}}c@{\hspace{0.3cm}}c@{\hspace{0.3cm}}c@{\hspace{0.3cm}}c@{\hspace{0.3cm}}c@{\hspace{0.3cm}}}
      \toprule[1.pt]
      \toprule[1.pt]
      \addlinespace[0.3em]
      & & \multicolumn{4}{c}{Median \bw/triangulation ratio} \\
      \addlinespace[0.2em]
      \cline{3-6}
      \addlinespace[0.2em]
      $m_1$ $[M_\odot]$ & $m_2$ $[M_\odot]$ & \multicolumn{2}{c}{50\% area} & \multicolumn{2}{c}{90\% area} \\
      & & incoherent & coherent & incoherent & coherent \\
      \addlinespace[0.2em]
      \midrule[1.pt]
      \addlinespace[0.4em]
      5   & 5   & 0.67 & 1.20 & 0.88 & 1.60 \\	
      5   & 10  & 0.62 & 1.12 & 1.06 & 2.06 \\
      10  & 10  & 0.54 & 1.05 & 0.74 & 1.32 \\
      5   & 50  & 0.42 & 0.88 & 0.46 & 0.97 \\
      10  & 50  & 0.38 & 0.76 & 0.36 & 0.75 \\
      50  & 50  & 0.32 & 0.78 & 0.32 & 0.84 \\
      5   & 100 & 0.21 & 0.62 & 0.25 & 0.71 \\
      10  & 100 & 0.25 & 0.70 & 0.28 & 0.73 \\
      50  & 100 & 0.19 & 0.61 & 0.23 & 0.71 \\
      100 & 100 & 0.13 & 0.47 & 0.19 & 0.75 \\
      \bottomrule[1.pt]
      \bottomrule[1.pt]
    \end{tabular}
  }
\end{table}

\subsection{Signal Classification}

The confident detection of unmodelled transients depends on the
ability to distinguish robustly true signals from the transient noise
fluctuations (``glitches'') that are common features of the detector
noise backgrounds~\cite{GW150914DetChar}.  Searches for generic
\ac{GW} transients typically rely on comparisons of weighted measures
of the cross-correlation between detectors to the total energy in the
data for signal-glitch discrimination (see, e.g.,
\cite{Chatterji:2006nh,X1,cWB2,TheLIGOScientific:2016uux}).  {\bw}
does this by calculating the log Bayes factors for the signal and
glitch hypotheses\footnote{The \texttt{oLIB} pipeline \cite{oLIB} 
performs a similar analysis, but restricted to a single wavelet.
For Bayesian signal-glitch discrimination that relies on the compact 
binary coalescences model, see e.g.~\cite{Veitch:2009hd,BCRmethod}).}.
Under each hypothesis, the transient (either the result of the two
\ac{GW} polarizations, or the glitch time-series in each detector) is
fit by a linear combination of Morlet-Gabor wavelets.  The Bayes
factor depends on both the quality of the fit and the priors;
generally, signals which have high \ac{SNR}-per-pixel throughout a
large time-frequency volume are most easily distinguished from
glitches.
Littenberg~\textit{et~al.}~\cite{Littenberg:2015kpb} argue that the
signal-vs.-glitch log Bayes factor $\log\mathcal{B}_{\mathcal{S},
  \mathcal{G}}$ can be approximated by
\begin{eqnarray}
  \log\mathcal{B}_{\mathcal{S}, \mathcal{G}} 
  & \simeq &  \frac{5N}{2} + 5N\log\left(\frac{\rho}{\sqrt{N}}\right) \nonumber\\
  &-&\sum_{n=1}^N \log\left(2^{13/6}\pi^{2/3} \mathcal{Q}_n\right)
  \nonumber\\
  & + & N \log V_\lambda + \left(2 + \log \frac{\sqrt{\det \mathcal{C}_Q}}{4 \pi^2} \right) \nonumber\\
  & \approx &  \frac{5N}{2} + 5N\log\left(\frac{\rho}{\sqrt{N}}\right) \, ,  \label{eq:bayes_SG}
\end{eqnarray}
where $N$ is the number of wavelets,
$\rho$ is the matched-filter \ac{SNR}, $V_{\lambda}$ is the volume of
the intrinsic parameter space, $\mathcal{C}_Q$ is the signal parameter
covariance matrix, and $\mathcal{Q}_n$ is the quality factor of the
$n^\mathrm{th}$ wavelet.  Equation (\ref{eq:bayes_SG}) is our
approximation, made by keeping only the leading order terms.

\begin{figure*}[htb]
  \begin{tabular*}{\textwidth}{c@{\extracolsep{\fill}}c}
    \includegraphics[width=0.5\textwidth,clip=true]{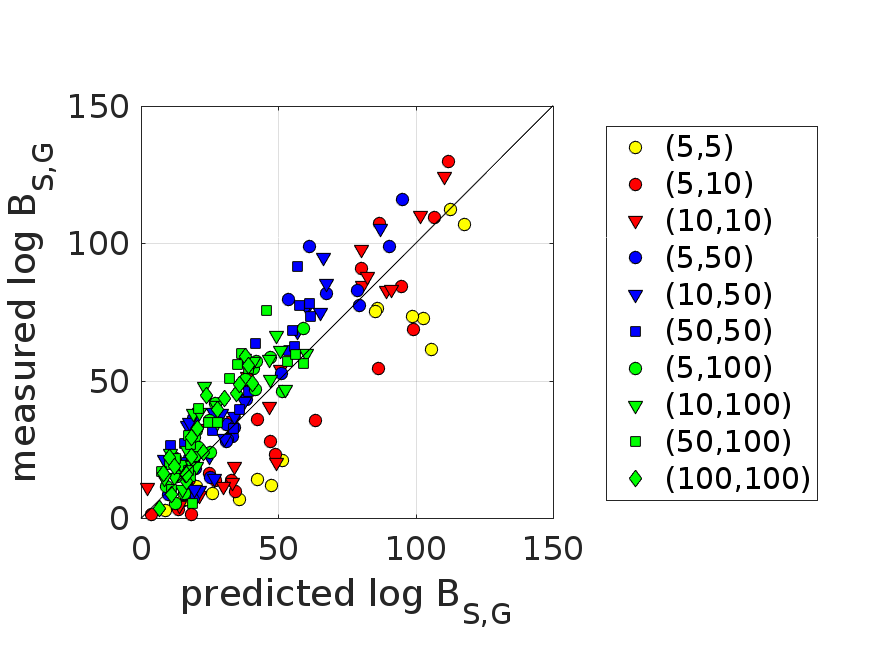}
    \includegraphics[width=0.5\textwidth,clip=true]{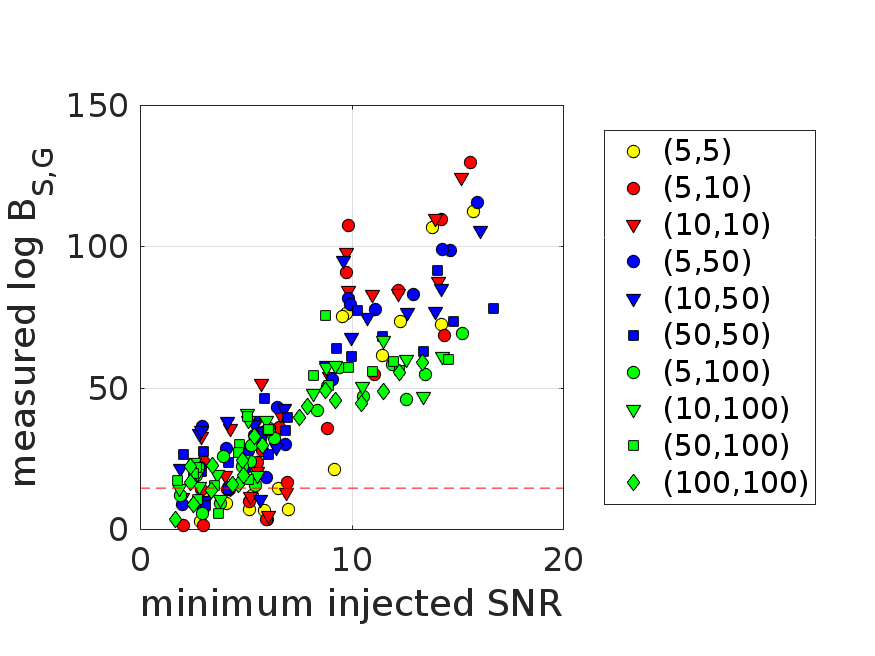}
  \end{tabular*}
  \caption{\label{fig:bayes_SG} (left) Predicted and measured log
    Bayes factors for the Signal~vs.~Glitch test. The measured log
    Bayes factors are in good agreement with the predicted analytical
    expressions from~\cite{Littenberg:2015kpb}
    and~\cite{Cornish:2014kda}, except for the lowest-mass systems,
    for which {\bw} is unable to recover the full \ac{SNR}.  (Right)
    Measured log Bayes factors vs.~minimum injected \ac{SNR}.  The red
    dashed line indicates a $\log\mathcal{B}_{\mathcal{S},
      \mathcal{G}}$ threshold that corresponds to a \ac{FAR} of
    1/100\,yr.  Low time-frequency volume signals are distinguishable
    from glitches at this \ac{FAR} provided the \ac{SNR} is greater
    than 5--6 in all three detectors.  High time-frequency volume
    signals are distinguishable for \acp{SNR} greater than 7--8 in all
    three detectors.}
\end{figure*}

We compare the analytical approximation of $\mathcal{B}_{\mathcal{S},
  \mathcal{G}}$ in Eq.\,(\ref{eq:bayes_SG}) with the measured \bw
output for a range of \ac{BBH} masses.  As discussed
in~\cite{Littenberg:2015kpb}, signal-glitch discrimination improves
with the number of detectors that see the transient.\footnote{A
  \ac{GW} can be fit with only two polarisations regardless of the
  number of detectors, while the glitch model needs to explain
  simultaneous independent noise fluctuations in each detector.}  We
find that the best predictions for $\mathcal{B}_{\mathcal{S},
  \mathcal{G}}$ come from using the \textit{minimum} injected \ac{SNR}
for $\rho$ in Eq.\,(\ref{eq:bayes_SG}); \textit{i.e.}, the lowest of
the \acp{SNR} in H1, L1 or V1.  The results are shown in the left
panel of Fig.\,\ref{fig:bayes_SG}, where the correlation between
measured and predicted $\log\mathcal{B}_{\mathcal{S}, \mathcal{G}}$ is
evident.  The measured log Bayes factors are lower than predicted for
the lowest-mass systems, because {\bw} is unable to recover the full
\ac{SNR} of these signals.  Low-mass systems require higher \ac{SNR}
per time-frequency pixel, which in turn limits their reconstruction
compared to high-mass systems.
The predictive power of Eq.\,(\ref{eq:bayes_SG}) can be improved
further by using the minimum \textit{recovered} \ac{SNR} instead of
the minimum injected \ac{SNR}.

We can compare these results to the typical log Bayes factor for
background noise to establish what real astrophysical signals could be
recovered with high confidence.  Using real LIGO noise from the
2009-10 run, Littenberg \textit{et al.}~\cite{Littenberg:2015kpb}
computed log Bayes factors for coincident events found by the \ac{cWB}
pipeline~\cite{cWB1, cWB2} and showed that a threshold of
$\log\mathcal{B}_{\mathcal{S}, \mathcal{G}} = 14.4$ corresponds to a
\ac{FAR} of 1/100\,yr.  In the first Advanced LIGO run, around the
time of GW150914, the same \ac{FAR} value corresponds to
$\log\mathcal{B}_{\mathcal{S}, \mathcal{G}} \sim 2$--$3$ (see Fig.\,4
in \cite{TheLIGOScientific:2016uux}).  For illustration, we use the
higher of these ($14.4$) as an indicative threshold; this is
represented by the red dashed line in the right panel of
Fig.\,\ref{fig:bayes_SG}.
We see that low time-frequency volume signals are distinguishable from
glitches at this \ac{FAR} provided the \ac{SNR} is greater than 5--6
in all three detectors, while high time-frequency volume signals are
distinguishable for \acp{SNR} greater than 7--8 in all three
detectors.

Finally, we note that {\bw} also provides a log Bayes factor for the
signal vs.~Gaussian noise hypotheses. Cornish and
Littenberg~\cite{Cornish:2014kda} show that this log Bayes factor can
be approximated by\footnote{Note that there is an error in Eq.\,(36)
  of \cite{Cornish:2014kda}: (1-FF$^2$) should be FF$^2$. We use the
  symbol $M$ (or match) instead of FF (fitting factor).}
\begin{eqnarray}
\log \mathcal{B}_{\mathcal{S}, \mathcal{N}} 
  & = &           M^2 \frac{\rho_\mathrm{net}^2}{2} + \Delta \ln \mathcal{O} \nonumber \\
  & \approx & M^2 \frac{\rho_\mathrm{net}^2}{2} \, , \label{eq:bayes_SN}
\end{eqnarray}
where $M$ is the match (discussed below) and $\rho_\mathrm{net}$ is
the coherent network \ac{SNR}. $\mathcal{O}$ is the Occam factor,
which we ignore for our comparisons.

\begin{figure}[htb]
  \includegraphics[width=1.1\columnwidth,clip=true]{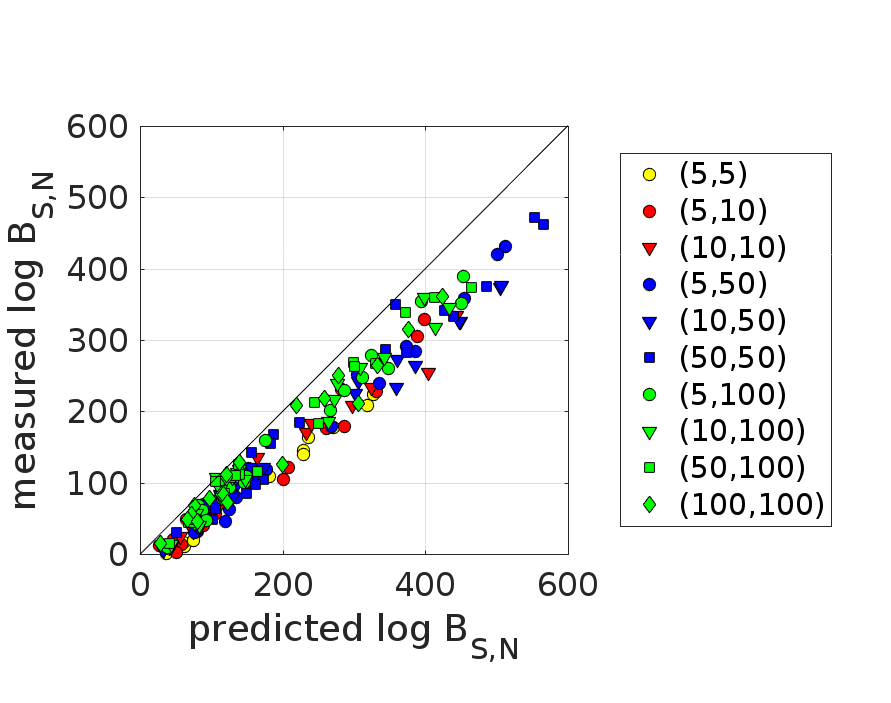}
  \caption{\label{fig:bayes_SN} Predicted and measured log Bayes
    factors for the Signal~vs.~Gaussian noise test. The measured log
    Bayes factors are in very good agreement with predicted analytical
    expressions from~\cite{Littenberg:2015kpb}
    and~\cite{Cornish:2014kda} based on the total injected network
    \ac{SNR} and the time-frequency volume of the signal. The measured
    log Bayes factors are systematically lower than the predicted ones
    for the lightest-mass (highest time-frequency volume) systems, for
    which {\bw} is unable to recover the full \ac{SNR}.}
\end{figure}

Figure \ref{fig:bayes_SN} shows that the predicted
$\log\mathcal{B}_{\mathcal{S}, \mathcal{N}}$'s closely follow the
measured $\log\mathcal{B}_{\mathcal{S}, \mathcal{N}}$'s.  As for
$\log\mathcal{B}_{\mathcal{S}, \mathcal{G}}$, we see that the measured
log Bayes factors are (slightly) lower than the predicted ones for the
lowest-mass systems, because {\bw} is unable to recover the full
\ac{SNR} of these signals.

\subsection{Waveform Reconstruction}
Many potential sources of \ac{GW} transients, such as core-collapse
supernov{\ae} and hypermassive neutron stars formed in binary neutron
star mergers, are too complicated to model accurately. In some cases
even parts of the underlying physics are unknown (\textit{e.g.}, the
neutron star equation of state). The ability to reconstruct the
received $h(t)$ signal without reliance on accurate ``templates'' will
therefore be crucial for the exploitation of \acp{GW} to probe new and
unexpected systems.

First-principles estimation of the match between the true \ac{GW}
signal with \ac{SNR} $\rho_\mathrm{inj}$ and a maximum-likelihood
reconstruction of the signal based on a time-frequency pixel analysis
can be estimated using only the recovered \ac{SNR} $\rho_\mathrm{rec}$
and number of pixels $N$~\cite{Sutton:inprep},
\begin{equation}
  \label{eqn:match}
  M  \simeq  \frac{\rho_\mathrm{rec}}{\rho_\mathrm{inj}} \left( 1 + \frac{2 N}{\rho_\mathrm{rec}^2} \right)^{-1/2}  
\end{equation}
with a one-sigma fractional uncertainty of
\begin{equation}
  \frac{\delta M}{M}  \simeq  \frac{\sqrt{3}}{\rho_\mathrm{rec}} \, .
\end{equation}
The $\rho_\mathrm{rec}/\rho_\mathrm{inj}$ factor in
Eq.\,(\ref{eqn:match}) is due to portions of the signal that are not
included in the reconstruction, such as the low-frequency early-time
portions of the low-mass signals.  The factor in parentheses is due to
the noise contamination of those pixels that are included in the
reconstructed waveform.  These expressions should be most accurate in
the limit of high \ac{SNR} per pixel, $\rho_\mathrm{rec}^2/N\gg1$.

\begin{figure}[!tb]
  \includegraphics[width=1.1\columnwidth,clip=true]{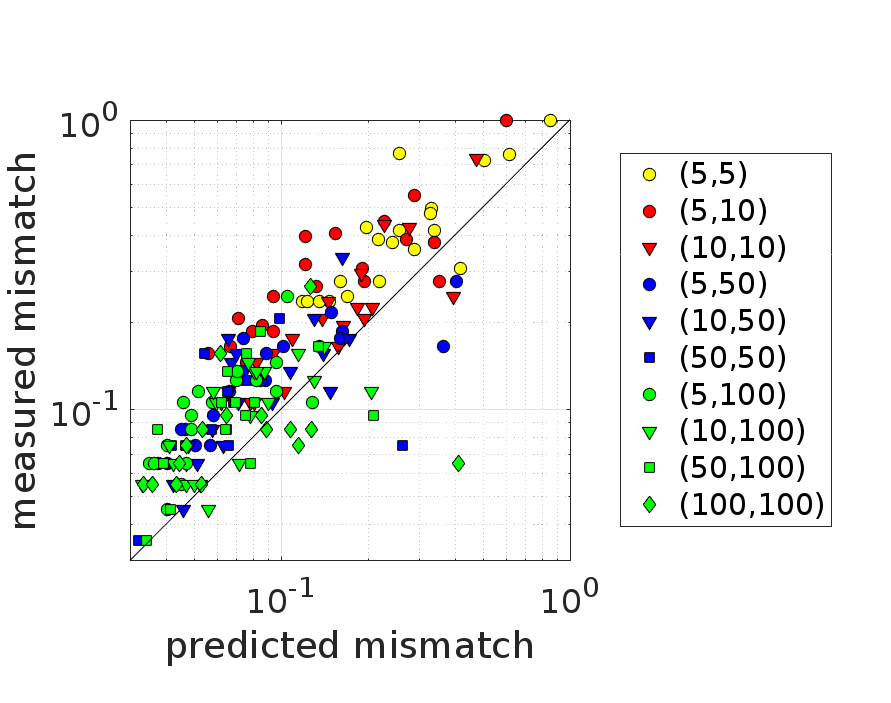}
  \caption{\label{fig:match} Measured mismatches between the true
    injected signal and that reconstructed by {\bw}, compared to a
    first-principles estimate of the lowest achievable mismatch.  The
    measured mismatches are in broad agreement with the
    first-principles estimate, but are typically 50\% higher.  The
    mismatches are smallest for the smallest time-frequency volume
    signals (high-mass systems), and largest for the largest
    time-frequency volume signals (low-mass systems) for which {\bw}
    is unable to reconstruct the full signal.  [For visual clarity, we
    do not show the error bars on the measured mismatches.]}
\end{figure}

Figure \ref{fig:match} compares the \textit{mismatch}, $1-M$, of the
waveform reconstructed by {\bw} to the first-principles estimate from
Eq.\,(\ref{eqn:match}). We see that there is broad agreement between
the two, with the measured mismatches typically about 50\% higher than
the first-principles estimate of the lowest achievable mismatch.  Not
surprisingly, the lowest mismatches are achieved for the signal with
smallest time-frequency volume (high masses), where the entire signal
in the sensitive band of the detectors is reconstructed.  The highest
mismatches are for the largest time-frequency volume signals (low
masses), where {\bw} is unable to reconstruct the full signal.  In
these cases, the mismatch is dominated by the {\bw} reconstruction not
including the full signal, as opposed to noise contamination of the
reconstruction.

\section{Summary and conclusions}
\label{sec:conclusions}
We have performed an in-depth analysis of the parameter estimation
capabilities of {\bw}, an algorithm for the reconstruction of \ac{GW}
signals without reference to a specific signal model.  Using simulated
\ac{BBH} signals added to simulated Advanced LIGO and Advanced Virgo
data, we evaluated {\bw} in three key areas: sky position estimation,
signal/glitch discrimination, and waveform reconstruction, comparing
its performance to first-principles estimates.  We found that {\bw}'s
effectiveness depends mainly on the time-frequency content of the
signal: the fewer wavelets needed to reconstruct a signal, the better
the performance.  Specifically, higher mass \ac{BBH} systems tend to
have shorter waveforms which can be accurately reconstructed with a
small number ($<10$) wavelets.

\bw localises the source on the sky better than timing-only
triangulation in all scenarios, and is outperformed by optimal
Bayesian matched-filter analyses only for low-mass systems ($\leq 50
\Msun$).  The measured log Bayes factor for signal-glitch
classification closely follows analytic predictions based on the
waveform match accuracy and the coherent network
\ac{SNR}. 
As a result, low time-frequency volume signals are distinguishable
from noise glitches provided \ac{SNR}$>5$--$6$ in all detectors, and
high time-frequency volume signals at \ac{SNR}$>7$--$8$, at a
false-alarm rate of 1/100y.  Finally, the match between reconstructed
and injected waveforms depends on the \ac{SNR} and the time-frequency
volume over which it is spread.  Low time-frequency volume signals can
achieve matches above 0.9, while high time-frequency volume signals
are more typically around a match of $0.6$--$0.8$.  The main
limitation for waveform reconstruction is the inability to reconstruct
the full signal when its \ac{SNR} is spread over a large number of
pixels, rather than noise contamination of the reconstructed signal.

While our study used \ac{BBH} signals,
a key strength of the {\bw} pipeline is that its performance does not
depend on signal morphology, so we expect to achieve similar results
for generic unmodelled \ac{GW} transients.
For example, it would be very interesting to assess the performance of
waveform reconstruction for signals from the post-merger remnant from
binary neutron star systems \cite{Abbott:2017dke}, given the recent
detection of GW170817~\cite{GW170817}.  Also, the waveforms used in
our study~\cite{Ajith:2009bn} do not include spin, eccentricity, or
higher-mode contributions for \ac{BBH} signals. While these effects
require substantial changes to waveform modelling and matched-filter
analyses, \bw should be able to account for all of these effects
automatically without modification.

{\bf \em Acknowledgements.}
This work was supported by STFC grants ST/L000962/1 and ST/N005430/1,
and by European Research Council Consolidator Grant 647839.  We wish
to thank Tyson Littenberg, Neil Cornish, and the LIGO-Virgo
collaboration Burst group for helpful discussions.  We also thank
Christopher Berry, Thomas Dent, Jonah Kanner, and Meg Millhouse for
useful comments on a draft of this work.

\providecommand{\newblock}{}

\bibliographystyle{iopart-num}

\end{document}